\RequirePackage[2020-02-02]{latexrelease}
\documentclass[aps, prm, twocolumn, amssymb, amsmath, float, superscriptaddress]{revtex4-1}
\usepackage{graphicx,dcolumn,bm,hyperref,subfigure,multirow,color,amsmath,systeme}
\begin{document}

\title{Machine-learning approach for discovery of conventional superconductors}

\author{Huan Tran}
\email{huan.tran@mse.gatech.edu}
\affiliation{School of Materials Science \& Engineering, Georgia Institute of Technology, 771 Ferst Dr. NW, Atlanta, GA 30332, USA}
\author{Tuoc N. Vu}
\affiliation{Institute of Engineering Physics, Hanoi University of Science \& Technology, 1 Dai Co Viet Rd., Hanoi 10000, Vietnam}

\begin{abstract}
	First-principles computations are the driving force behind numerous discoveries of hydride-based superconductors, mostly at high pressures, during the last decade. Machine-learning (ML) approaches can further accelerate the future discoveries if their reliability can be improved. The main challenge of current ML approaches, typically aiming at predicting the critical temperature $T_{\rm c}$ of a solid from its chemical composition and target pressure, is that the correlations to be learned are deeply hidden, indirect, and uncertain. In this work, we showed that predicting superconductivity at any pressure from the atomic structure is sustainable and reliable. For a demonstration, we curated a diverse dataset of 584 atomic structures for which $\lambda$ and $\omega_{\log}$, two parameters of the electron-phonon interactions, were computed. We then trained some ML models to predict $\lambda$ and $\omega_{\log}$, from which $T_{\rm c}$ can be computed in a post-processing manner. The models were validated and used to identify two possible superconductors whose $T_{\rm c}\simeq 10-15$K at zero pressure. Interestingly, these materials have been synthesized and studied in some other contexts. In summary, the proposed ML approach enables a pathway to directly transfer what can be learned from the high-pressure atomic-level details that correlate with high-$T_{\rm c}$ superconductivity to zero pressure. Going forward, this strategy will be improved to better contribute to the discoveries of new superconductors.
\end{abstract}

\maketitle

\section{Introduction}

In the search for high critical temperature ($T_{\rm c}$) superconductors, significant progress has been made during the last decade \cite{Drozdov15, drozdov2019superconductivity, snider2020room}. Among thousands of hydride-based superconducting materials computationally predicted \cite{duan2014pressure, zurek2019high, hilleke2022tuning,errea2020quantum, gao2021superconducting, boeri20212021, yazdani2022artificial, Alexey:CaB, zhang2022design}, mostly at very high pressures, e.g., $P \gtrsim 100$ GPa, dozens of them, e.g., H$_3$S \cite{Drozdov15}, LaH$_{10}$ \cite{drozdov2019superconductivity}, and CSH \cite{snider2020room}, were synthesized and tested. This active research area is presumably motivated by Ashcroft, who, in 2004, predicted \cite{ashcroft2004hydrogen} that high-$T_{\rm c}$ superconductivity may be found in hydrogen dominant metallic alloys, probably at high $P$. Another driving force is the development of first-principles computational methods to predict material structures at any $P$ \cite{OganovBook, oganov2019structure, AIRSS, NeedsReview:CMD, Huan:hafnia, Huan:Mg-Si, Huan:perovskites} and to calculate the electron-phonon (EP) interactions \cite{DFPT,giustino2017electron}, the atomic mechanism behind the conventional superconductivity, according to the Bardeen-Cooper-Schrieffer (BCS) theory \cite{bardeen1957microscopic}. While critical debates on some discoveries \cite{hirsch2021unusual, wang2021absence, hirsch2021nonstandard, gubler2022missing, eremets2022high, hirsch2022superconducting} are on-going, it seems that the one-day-realized dream of superconductors at ambient conditions may be possible. Readers are referred to some reviews \cite{zurek2019high, hilleke2022tuning,gao2021superconducting, pickett2022room} and a recent roadmap \cite{boeri20212021} for progresses, challenges, and future pathways of this research area.

\begin{figure}[b]
\centering
\includegraphics[width=1.0\linewidth]{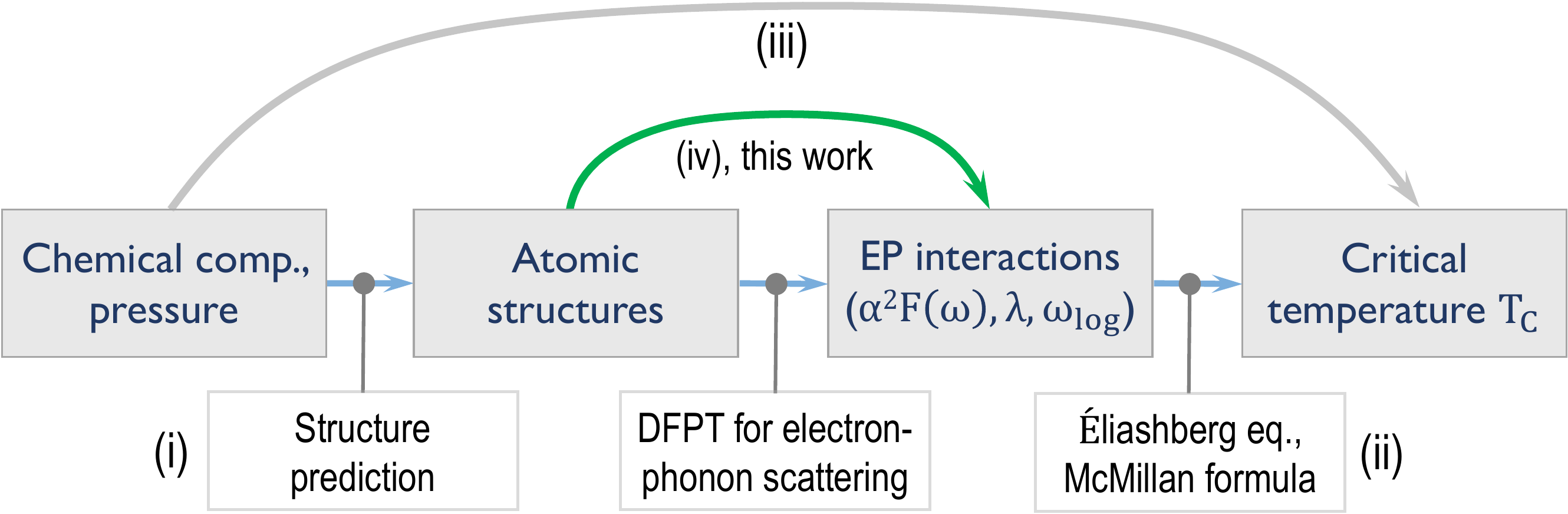}
\caption{A typical workflow to compute $T_{\rm c}$. Existing ML efforts are devoted to (i) using ML potentials to accelerate the structure prediction step, (ii) deriving new formulas of $T_{\rm c}$, and (iii) predicting $T_{\rm c}$ from chemical composition. This work is in (iv), predicting $\lambda$ and $\omega_{\log}$ from atomic structures.}\label{fig:flow}
\end{figure}

The central role of first-principles computations in the recent discoveries of conventional superconductors stems from \'{E}liashberg theory \cite{Eliashberg, allen1983theory, marsiglio2020eliashberg, chubukov2020eliashberg}, of which the spectral function $\alpha^2F(\omega)$ characterizing the EP interactions could be evaluated numerically. The first inverse moment $\lambda$ and logarithmic moment $\omega_{\log}$ of $\alpha^2F(\omega)$, together with an empirical Coulomb pseudopotential $\mu^*$, are the inputs to estimate $T_{\rm c}$ by either solving the \'{E}liashberg equations \cite{Eliashberg, allen1983theory, marsiglio2020eliashberg, chubukov2020eliashberg} or using the McMillan formula \cite{McMillanTc, dynes1972mcmillan, AllenTc} (see Sec. \ref{sec:theory} for more details). In a typical workflow (Fig. \ref{fig:flow}), a search for stable atomic structures across multiple related chemical compositions is performed at a given pressure, usually with first-principles computations. Then, $\alpha^2F(\omega)$, $\lambda$, $\omega_{\log}$, and finally $T_{\rm c}$ are evaluated, identifying candidates with high estimated $T_{\rm c}$ for possible new superconducting materials. Although structure prediction \cite{OganovBook, oganov2019structure, AIRSS, NeedsReview:CMD} and $\alpha^2F(\omega)$ computations \cite{DFPT,giustino2017electron} are extremely expensive and technically non-trivial, significant research efforts have been devoted to and shaped by this workflow.

Machine-learning (ML) methods have recently emerged in the discoveries of superconductors \cite{yazdani2022artificial,boeri20212021}. As sketched in Fig. \ref{fig:flow}, existing ML efforts can be categorized into four lines, including (i) using some ML potentials to accelerate the structure prediction step \cite{yang2021hard}, (ii) using some symbolic ML techniques to derive new empirical expressions for $T_{\rm c}$ \cite{xie2019functional, xie2022machine}, (iii) developing some ML models to predict $T_{\rm c}$ from a chemical composition at a given pressure $P$ \cite{hamidieh2018data, matsumoto2019acceleration, ishikawa2019materials, shipley2021high, song2021high, le2020critical, garcia2021prediction, stanev2021artificial, raviprasad2022tree, revathy2022random}, and (iv) developing some ML models to predict $\lambda$, $\omega_{\log}$, and $\alpha^2F(\omega)$ from the atomic structures \cite{choudhary2022designing}. While line (iii) is predominant, its role remains limited, presumably because the connections from the chemical composition and the target $P$ to $T_{\rm c}$ are deeply hidden. In fact, there are at least two ``missing links'' between the two ends of this approach. One of them is the atomic-level information while the other is the microscopic mechanism of the superconductivity, e.g., the EP interactions in conventional superconductors. The former is critical because for a given chemical composition, the properties of thermodynamically competing atomic structures can often be fundamentally different, e.g., one is insulating and another is conducting \cite{HuanData2021, Huan:hafnia}. Therefore, ignoring the atomic structure is equivalent to adding an irreducible uncertainty into the ML predictions \cite{Tuoc:PrNN}. Likewise, the latter cannot be overemphasized. In fact, bypassing $\alpha^2F(\omega)$, $\lambda$, and $\omega_{\log}$, and using an empirical value of $\mu^*$ are intractable assumptions, and thus, uncontrollable approximations. In line (iv), initialized recently by Ref. \onlinecite{choudhary2022designing} during the (independent) preparation of this work, these missing links are addressed in some ways.

In this paper, we present an initial step to bring the atomic-level information into the ML-driven pathways toward new conventional (or BCS) superconductors, especially at ambient pressure. For this goal, we curated a dataset of 584 atomic structures for which more than 1,100 values of $\lambda$ and $\omega_{\log}$ were computed at different values of $P$ and reported, mostly in the last decade. The obtained dataset was visualized, validated, and standardized before being used to develop ML models for $\lambda$ and $\omega_{\log}$. Then, they were used to screen over $80,000$ entries of Materials Project database \cite{Jain2013}, identifying and confirming (by first-principles computations) two thermodynamically and dynamically stable materials whose superconductivity may exist at $T_{\rm c} \simeq 10-15$K and $P = 0$. We also proposed a procedure to compute $\lambda$ and $\omega_{\log}$, for which convergence are generally hard to attain \cite{choudhary2022designing}. 

This scheme relies on the direct connection between the atomic structures and $\lambda$ and $\omega_{\log}$, quantitatively described in Sec. \ref{sec:theory}. Pressure is an {\it implicit} input, i.e., $P$ determines the atomic structures for which $\lambda$ and $\omega_{\log}$ are computed/predicted. The design of this scheme has some implications. First, the ML models are trained on the atomic structures realized at high $P$ and (computationally) proved to correlate with high-$T_{\rm c}$ superconductivity. These structures can be considered ``unusual'' in the sense that their high-$P$ atomic-level details, e.g., short bond lengths and distorted bond angles, are not usually realized at zero pressure. Therefore, we hope that the ML models can identify the atomic structures realized at $P=0$ with relevant unusual atomic-level features, and thus, they may exhibit possible high-$T_{\rm c}$ superconductivity. Second, massive material databases \cite{choudhary2022recent} like Materials Project \cite{Jain2013}, OQMD \cite{OQMD} and NOMAD with millions of atomic structures can now be screened directly with robust and reliable ML models. Given that only a small search space was explored in this demonstrative work, we expect more superconducting materials to be discovered in the next steps of our effort.

\section{Methods}
\subsection{\'{E}liashberg theory and McMillan formula}\label{sec:theory}
In \'{E}liashberg theory \cite{Eliashberg, allen1983theory, marsiglio2020eliashberg, chubukov2020eliashberg}, $\alpha^2F(\omega)$ is a spectral function characterizing the EP scattering, which is defined as
\begin{equation}\label{eq:a2f}
\alpha^2F(\omega) = \frac{1}{N_0}\sum_{{\bf kk'}ij\nu} \vert g^{ij,\nu}_{\bf k, k'}\vert^2 \delta(\varepsilon^i_{\bf k})\delta(\varepsilon^j_{\bf k'}) \delta(\omega-\omega^\nu_{\bf k-k'}).
\end{equation}
Here, $N_0$ is the density of states at the Fermi level, $g^{ij,\nu}_{\bf k, k'}$ the electron-phonon matrix elements, $\nu$ the polarization index of the phonon with frequency $\omega$, $\delta$ the delta-Dirac function, and (${\bf k}$ and ${\bf k'}$)/($\varepsilon^i_{\bf k}$ and $\varepsilon^j_{\bf k'}$) the (electron wave vectors)/(band energies) corresponding to the band indices ($i$ and $j$), respectively. 

The standard method to compute $\alpha^2F(\omega)$ is density functional perturbation theory (DFPT) \cite{DFPT,giustino2017electron}, as implemented in major codes like Quantum ESPRESSO \cite{QE1, QE2} and ABINIT \cite{Gonze_Abinit_1,Gonze_Abinit_2,Gonze_Abinit_3}. Having $\alpha^2F(\omega)$, $T_{\rm c}$ can be evaluated by numerically solving a set of (unfortunately, quite complicated) \'{E}liashberg equations using, for example, Electron-Phonon Wannier (EPW) \cite{giustino2007electron,margine2013anisotropic, ponce2016epw}. The much more frequent method to estimate $T_{\rm c}$ is using some empirical formulas derived from the \'{E}liashberg equations. Perhaps the most extensively used formula is
\begin{equation}\label{ADM0}
T_{\rm c} = \frac{\omega_{\log}}{1.2}\exp\left[-\frac{1.04(1+\lambda)}{\lambda-\mu^*(1+0.62\lambda)}\right],
\end{equation}
which was developed by McMillan \cite{McMillanTc} and latter improved by Allen and Dynes \cite{dynes1972mcmillan, AllenTc}. Here, 
\begin{equation}\label{eq:lambda}
	\lambda = 2\int_0^\infty d\omega\frac{\alpha^2F(\omega)}{\omega}
\end{equation}
is the (averaged) isotropic EP coupling while 
\begin{equation}\label{eq:omlog}
	\omega_{\log} = \exp\left[\frac{2}{\lambda}\int_0^\infty d\omega\ln(\omega)\frac{\alpha^2F(\omega)}{\omega}\right]. 
\end{equation}
Following Ashcroft \cite{ashcroft2004hydrogen}, the Coulomb pseudopotential $\mu^*$, which appears in Eq. \ref{ADM0} and connects with $N_0$, was empirically chosen in the range between 0.10 and 0.15. Eq. (\ref{ADM0}) indicates that in general, high values of $\lambda$ and/or $\omega_{\rm log}$ are needed for a high value of $T_{\rm c}$. Some new empirical formulas of $T_{\rm c}$ were developed recently \cite{xie2019functional, xie2022machine} using some symbolic ML techniques. Moving forward, developing a truly {\it ab-initio} framework for computing $T_{\rm c}$ \cite{luders2005ab, marques2005ab, sanna2020combining} is desirable and currently active.

The McMillan formula (\ref{ADM0}) is believed to be good for $\lambda\leq 1.5$ while additional empirical parameters are needed for larger $\lambda$ \cite{AllenTc}. Nevertheless, the exponential factor of Eq. \ref{ADM0} has a singular point at $\lambda = \mu^*/(1-0.62\mu^*)$, which could lead to unwanted/unphysical divergence. If we select $\mu^* = 0.1$ (or 0.15), $T_{\rm c}\to\infty$ when $\lambda$ approaches $0.1066$ (or $0.1654$) from below. Such values of $\lambda$ have been realized in many computational works \cite{xie2014superconductivity, kim2009predicted, di2020phase, xie2020hydrogen}, although much larger values, e.g., $\lambda \geq 0.7$, are generally needed for high-$T_{\rm c}$ superconductors. Given these observations, we believe that a ML approach for discovering conventional superconductor should focus on $\lambda$, $\omega_{\rm log}$, and perhaps $\alpha^2F(\omega)$, from which $T_{\rm c}$ can easily be estimated using, for examples, Eq. (\ref{ADM0}).

\subsection{Basic idea and approach}
The ML approach used in this work focuses on predicting $\lambda$ and $\omega_{\log}$ from the atomic structure of the considered materials. As visualized in Fig. \ref{fig:flow}, the role of $P$ is embedded in the main input of this scheme, i.e., the atomic structure, which is determined from $P$. The rationale of this design is two fold. First, what this ML approach will learn is a {\it direct and physics-inspired} correlation from an atomic structure to $\lambda$ and $\omega_{\log}$ through $\alpha^2F(\omega)$, as quantitatively described in Eqs. (\ref{eq:a2f}), (\ref{eq:lambda}), and (\ref{eq:omlog}). Second, the training data, which include the atomic environments/structures realized at multiple values of (sometimes very high) pressure $P$ that could lead to very high values of $\lambda$ and $\omega_{\log}$, will be highly diverse and comprehensive. Consequently, the resulted ML models will thus be robust, reliable, and, more importantly, they can be used to recognize new high-$T_{\rm c}$ superconductors that resemble unusual atomic-level details {\it at any pressure}, specifically $P=0$ GPa. This approach involves some challenges, one of them is how to obtain good datasets for the learning scheme. Our solution is described below.

\subsection{Data curation}\label{sec:data}
This work requires a dataset of the atomic structures for which $\lambda$, $\omega_{\log}$, and $\alpha^2F(\omega)$ were computed and reported. The curation of such a dataset is painstaking. Scientific articles published during the last $10-15$ years, reporting computed superconducting properties of new or known materials, were collected. In majority of the articles, the atomic structures were reported in some Tables while electronic files of standard formats, e.g., crystallographic information file (CIF), were given in very few cases. In some cases, important information, e.g., angle $\beta$ in a monoclinic structure, was missing from the Tables. When the provided information is sufficient, we used the obtained crystal symmetry/space group, lattice parameters, Wyckoff positions, and the coordinates of the inequivalent atoms, to manually reconstruct the reported structures. All the atomic structures obtained from electronic files and/or reconstructed from data Tables were inspected visually. During this step, a good number of them were found to be clearly incorrect, largely because of typos, number overrounding, and other possible unidentified reasons, when reporting the data. Incorrect structures were discarded.

\begin{figure}[t]
\centering
	\includegraphics[width=1.0\linewidth]{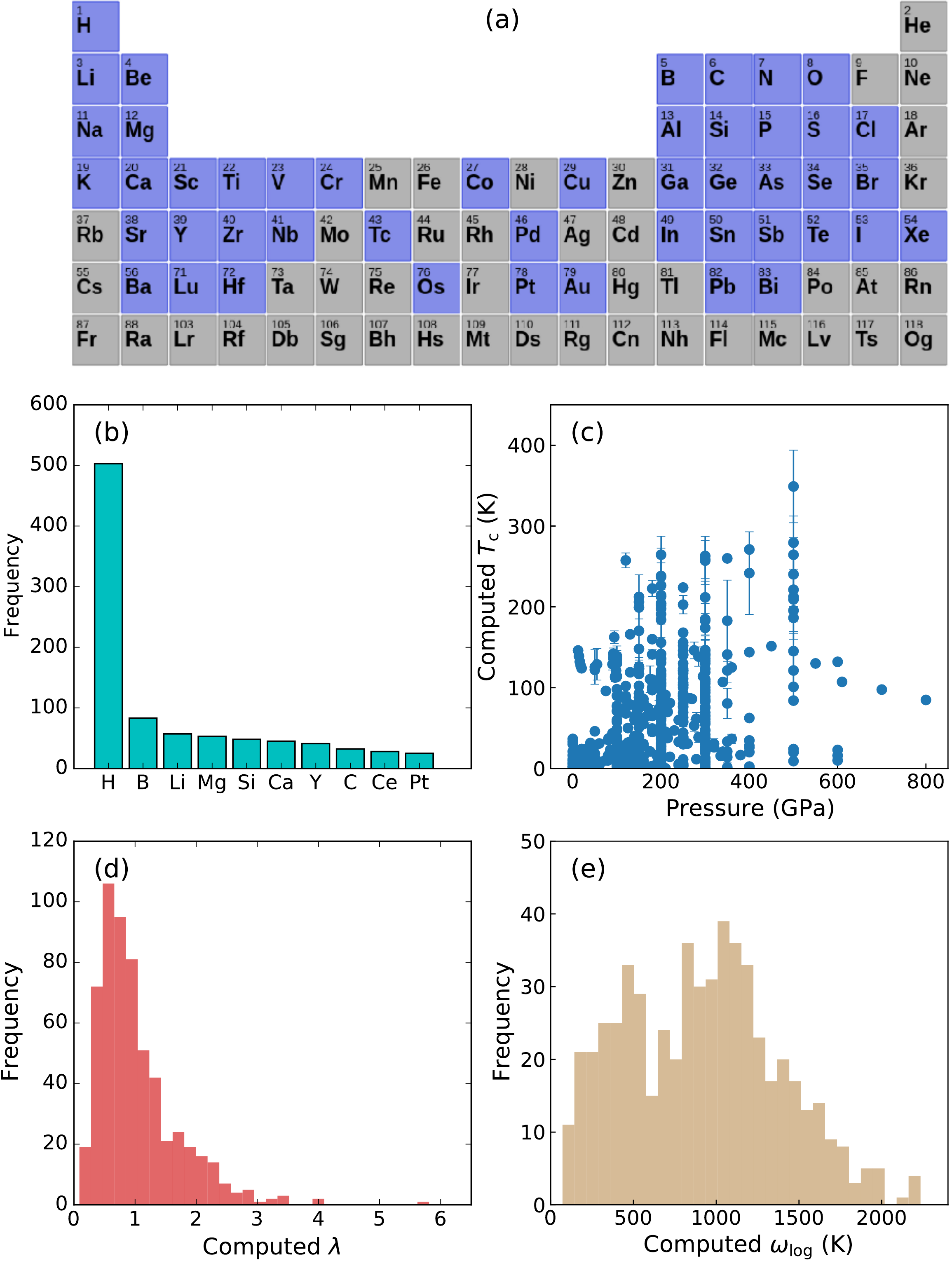}
	\caption{A summary of computed $\lambda$, $\omega_{\log}$, and $T_{\rm c}$ dataset of 584 superconducting materials reported and curated, including (a) the Periodic Table coverage, (b) 10 most-frequent species in the dataset, (c) 567 values of $T_{\rm c}$ computed and arranged at different pressures, and the distribution of (d) 584 computed values of $\lambda$, and (e) 567 values of $\omega_{\log}$, are given. Among 53 species found in our dataset, 47 of them are shown in (a) and the other 6 species are Ac, Ce, La, Nd, Pm, and Pr. In (b), each solid circle represents a combination of a chemical composition and a pressure while errorbars are for cases $T_{\rm c}$ was computed for different atomic structures, using different methods, e.g., using McMillan formula and solving \'{E}liashberg equations, and/or different values of $\mu^*$.}\label{fig:data}
\end{figure}

Superconducting-related properties, e.g., $\lambda$, $\omega_{\log}$, and $T_{\rm c}$, which were computed and reported for the atomic structures at pressure $P$ up to 800 GPa, were collected. These properties were mainly computed by some major workhorses like Quantum ESPRESSO \cite{QE1, QE2}, ABINIT \cite{Gonze_Abinit_1,Gonze_Abinit_2,Gonze_Abinit_3}, and EPW codes \cite{giustino2007electron,margine2013anisotropic, ponce2016epw}, employing different pseudopotentials, XC  functionals, energy cutoffs, smearing width needed to compute the $\delta$ functions appearing in the expression (\ref{eq:a2f}) of $\alpha^2F(\omega)$, and more. We recognize that data of $\lambda$ and $\omega_{\log}$ curated from scientific literature are not entirely uniform; they rather contain a certain level of uncertainty that will inevitably be translated into the (aleatoric) uncertainty of the predictions \cite{Tuoc:PrNN}. However, the demonstrated reproducibility of advanced first-principles computations \cite{lejaeghere2016reproducibility} suggests that data carefully produced by major codes should still be consistent and reliable.

To further improve the uniformity of the data, we used density functional theory (DFT) \cite{DFT1, DFT2} calculations to optimize the obtained atomic structures at the pressures reported, employing the same technical details used for Materials Project database. The rationale behind this step is that the predictive ML models trained on the dataset will then be used to predict $\lambda$ and $\omega_{\log}$ for the atomic structures obtained from Materials Project database. Therefore, the training data should be prepared at the same level of computations with the input data for predictions. In fact, a vast majority of our DFT optimizations were terminated after about a dozen steps or below, indicating that they were already optimized very well. Details of the optimizations are given in Sec. \ref{sec:dft}. Compared with the DFPT calculations for $\lambda$ and $\omega_{\log}$, the optimization step is computationally negligible. 

Our dataset includes 584 atomic structures for which at least $\lambda$ was computed and reported. Among them, 567 atomic structures underwent $\omega_{\log}$ and thus, $T_{\rm c}$ calculations (there is a trend in the community that computed $\lambda$ is more likely to be reported than $\omega_{\log}$ when discussing the superconductivity). Our dataset, which is summarized in Figs. \ref{fig:data} (a), (b), (c), (d), and (e), contains 53 species and covers a substantial part of the Periodic Table. Five most frequently encountered species are H, B, Li, Mg, and Si, which were found in 505, 83, 57, 53, and 48 entries, respectively. The dominance of H in this dataset reflects the focus of the community on super hydrides when searching for high-$T_{\rm c}$ superconductors. For $\lambda$, the smallest value is $0.089$, reported in Ref. \onlinecite{xie2014superconductivity} for the $P4/mbm$ structure of LiH$_2$ at $P = 150$ GPa while the largest value is $5.81$ reported in Ref. \onlinecite{shipley2021high} for the $Im\overline{3}m$ structure of CaH$_6$ at $P = 100$ GPa. Likewise, the smallest value of $\omega_{\rm log}$ is $71$ K reported in Ref. \onlinecite{zhang2020high} for the $I4/mmm$ structure of TiH at $P=50$ GPa whereas the largest value is 2,234 K reported in Ref. \onlinecite{shipley2021high} for the $P\overline{6}2m$ structure of CaH$_{15}$ at $P = 500$ GPa. Figs. \ref{fig:data} (d) and (e) provide two histograms summarizing the $\lambda$ and $\omega_{\log}$ datasets.

\subsection{Data representation and machine-learning approaches}
Materials atomic structures are not naturally ready for ML algorithms. The main reason is that they are not invariant with respect to transformations that do not change the materials in any physical and/or chemical ways, e.g., translations, rotations, and permutations of alike atoms. Therefore, we used {\sc matminer} \cite{WARD201860}, a package that offers a rich variety of material features, to convert (or featurize) the atomic structures into numerical vectors, which meet the requirements of invariance and can be used to train ML models. Starting from several hundreds components, optimal sets of features (the vector components) were determined using the recursive feature elimination algorithm as implemented in {\sc scikit-learn} library \cite{scikit-learn}. The final version of the $\lambda$ and $\omega_{\rm log}$ datasets have 40 and 38 features, respectively.

In principle, two featurized datasets of $\lambda$ and $\omega_{\log}$ can be learned simultaneously using a multi-task learning scheme so that the underlying correlations between $\lambda$ and $\omega_{\log}$ may be exploited. However, the intrinsically deep correlations in materials properties require a sufficiently big volume of data to be revealed. We have tested some multi-task learning schemes and found that with a few hundreads data points, they are not significantly better than learning $\lambda$ and $\omega_{\log}$ separately. In fact, similar behaviors are commonly observed in the literature \cite{Tuoc:PrNN}. Therefore, we examined six typical ML algorithms, including support vector regression, random forest regression, kernel ridge regression, Gaussian process regression, gradient boosting regression, and artificial neural networks two develop ML models for $\lambda$ and $\omega_{\log}$. For each algorithm, we created a pair of learning curves and used them to analyze the performance of the algorithm on the data we have. By carefully tuning the possible model parameters and examining the training and the validation curves, Gaussian process regression (GPR) \cite{GPR95, GPRBook} was selected. Details on the learning curves and the GPR models used for predicting $\lambda$ and $\omega_{\log}$ are discussed in Sec. \ref{sec:model}.

\subsection{First-principles calculations}\label{sec:dft}
First-principles calculations are needed for two purposes, i.e., to uniformly optimize the curated atomic structures and to compute $\alpha^2F(\omega)$, $\lambda$, and $\omega_{\log}$ for those identified by the ML models we developed. For the first objective, we followed the technical details used for Materials Project database, employing {\sc vasp} code \cite{vasp1, vasp3}, the standard PAW pseudopotentials, a basis set of plane waves with kinetic energy up to 520 eV, and the generalized gradient approximation Perdew-Burke-Ernzerhof (PBE) exchange-correlation (XC) functional. \cite{PBE} Convergence in optimizing the structures was assumed when the atomic forces become $< 10^{-2}$ eV/\AA ~after no more than 3 iterations.

In the computations of $\alpha^2F(\omega)$, $\lambda$, and $\omega_{\log}$, we used the version of DFPT implemented in ABINIT package \cite{Gonze_Abinit_1,Gonze_Abinit_2,Gonze_Abinit_3}, which also offers a rich variety of other DFT-based functionalities. Within this numerical scheme, we used the optimized norm-conserving Vanderbilt pseudopotentials (ONCVPSP-PBE-PDv0.4) \cite{hamann2013optimized} obtained from the PseudoDojo library \cite{van2018pseudodojo} and the PBE XC functional \cite{PBE}. The kinetic energy cutoff we used is 60 Hatree ($\simeq 1,600$eV), which is twice larger than the value suggested \cite{hamann2013optimized} for these norm-conserving pseudopotentials. The smearing width for computing $\alpha^2F(\omega)$ is $5\times 10^{-6}$ Ha, i.e., $\simeq 0.032$ THz. This value was selected to be $< 0.1$\% of the entire range of frequency while covering more than 4 (numerical) spacings of the frequency grid. 

Before entering the electron-phonon calculations with DFPT, the material structures under consideration were {\it repeatedly} optimized until the maximum atomic force is below $10^{-5}$ Hatree/bohr, which is $\simeq 5.1\times 10^{-4}$ eV/\AA, after no more than 3 iterations. Because the optimizations need the simulation box to change its shape, such a small number of iterations is required to minimize the cell volume change, thereby limiting the Pulay stress, and ultimately ensuring an absolute convergence of the force calculations. This level of accuracy is generally needed for phonon-related calculations.

\begin{figure}[t]
\centering
\includegraphics[width=0.95\linewidth]{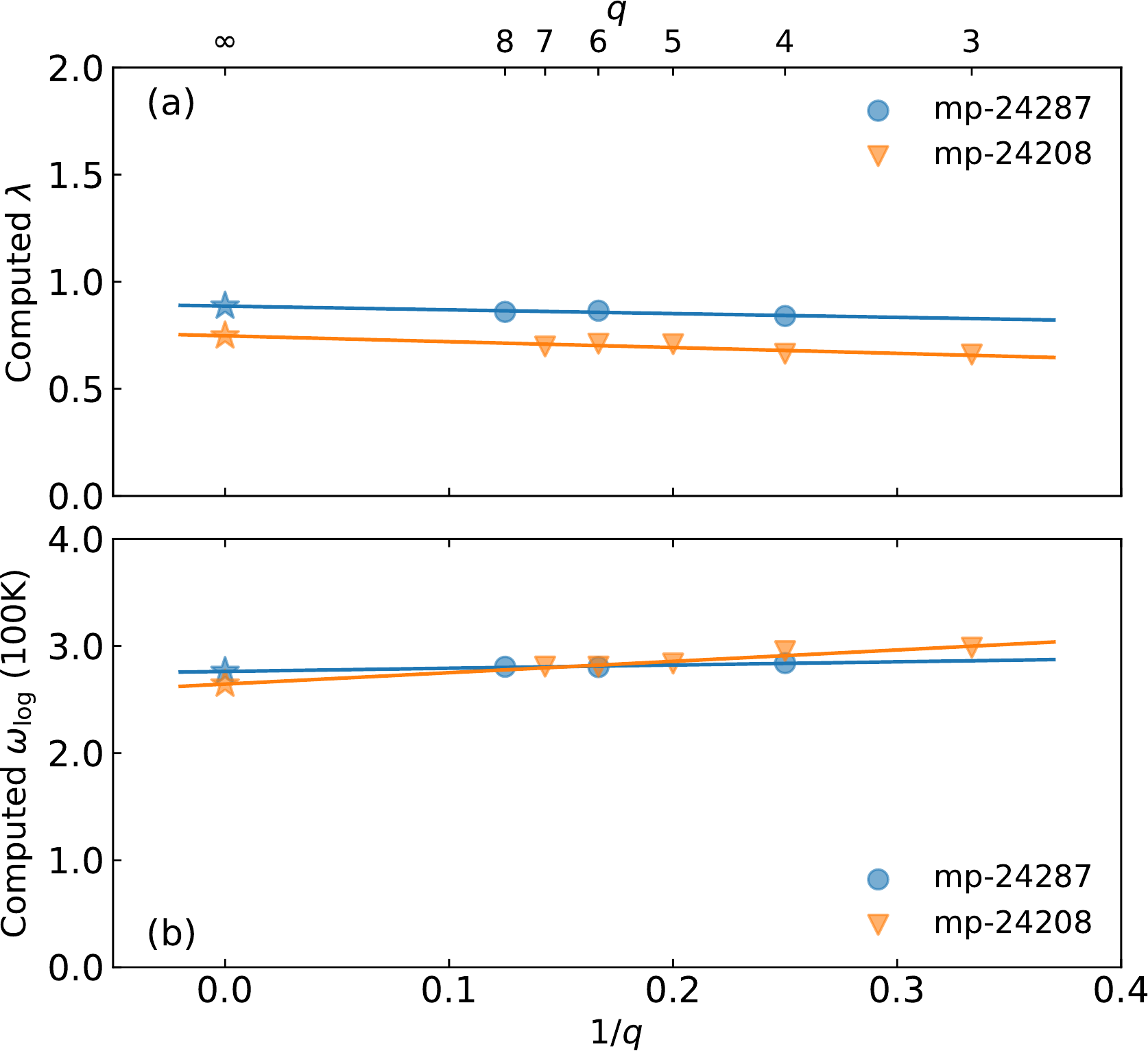}
\caption{Fitting procedure used to compute (a) $\lambda$ and (b) $\omega_{\log}$ of mp-24287 and mp-24208, two atomic structures identified from Materials Project database. Solid symbols show $\lambda$ and $\omega_{\log}$ computed with some finite $\bf q$-point grids while stars represent the extrapolated values of $\lambda$ and $\omega_{\log}$ at the limit of infinite $\bf q$-point grid, i.e., $1/q = 0$.}\label{fig:fitting}
\end{figure}

Eq. \ref{eq:a2f} indicates that $\alpha^2F(\omega)$ is evaluated on a $\bf q$-point grid of ${\bf q = k-k'}$, which must be a sub-grid of the full {\bf k}-point grid used to sample the Brillouin zone for regular DFT calculations. Therefore, calculations of $\alpha^2F(\omega)$ are extremely heavy while the convergence with respect to the $\bf q$-point grid is critical and must be examined \cite{shipley2021high, choudhary2022designing}. For this goal, we first computed $\alpha^2F(\omega)$, $\lambda$, and $\omega_{\log}$ using several $\bf q$-point grids of $q\times q\times q$ and $\bf k$-point grids of $k\times k\times k$ where $q$ is as large as possible depending on the structure size and $k \geq 3\times q$. Then, the computed values of $\lambda$ and $\omega_{\log}$ are fitted to a linear function of $1/q$. The values of the fitted functions at $1/q = 0$, or, equivalently, at the limit of $q\to \infty$, are the values assumed for $\lambda$ and $\omega_{\log}$. This procedure is visualized in Fig. \ref{fig:fitting} when $\lambda$ and $\omega_{\log}$ of two atomic structures reported in this work were computed. Details on the $\bf q$-point and $\bf k$-point grids and the corresponding computed data used for the fitting procedure can be found in Supplemental Material \cite{supplement}. A technique of similar philosophy has been demonstrated \cite{tran2022toward} in the computations of ring-opening enthalpy, the thermodynamic quantity that controls the ring-opening polymerizations.

\subsection{Candidates}
\begin{figure}[t]
\centering
\includegraphics[width=0.9\linewidth]{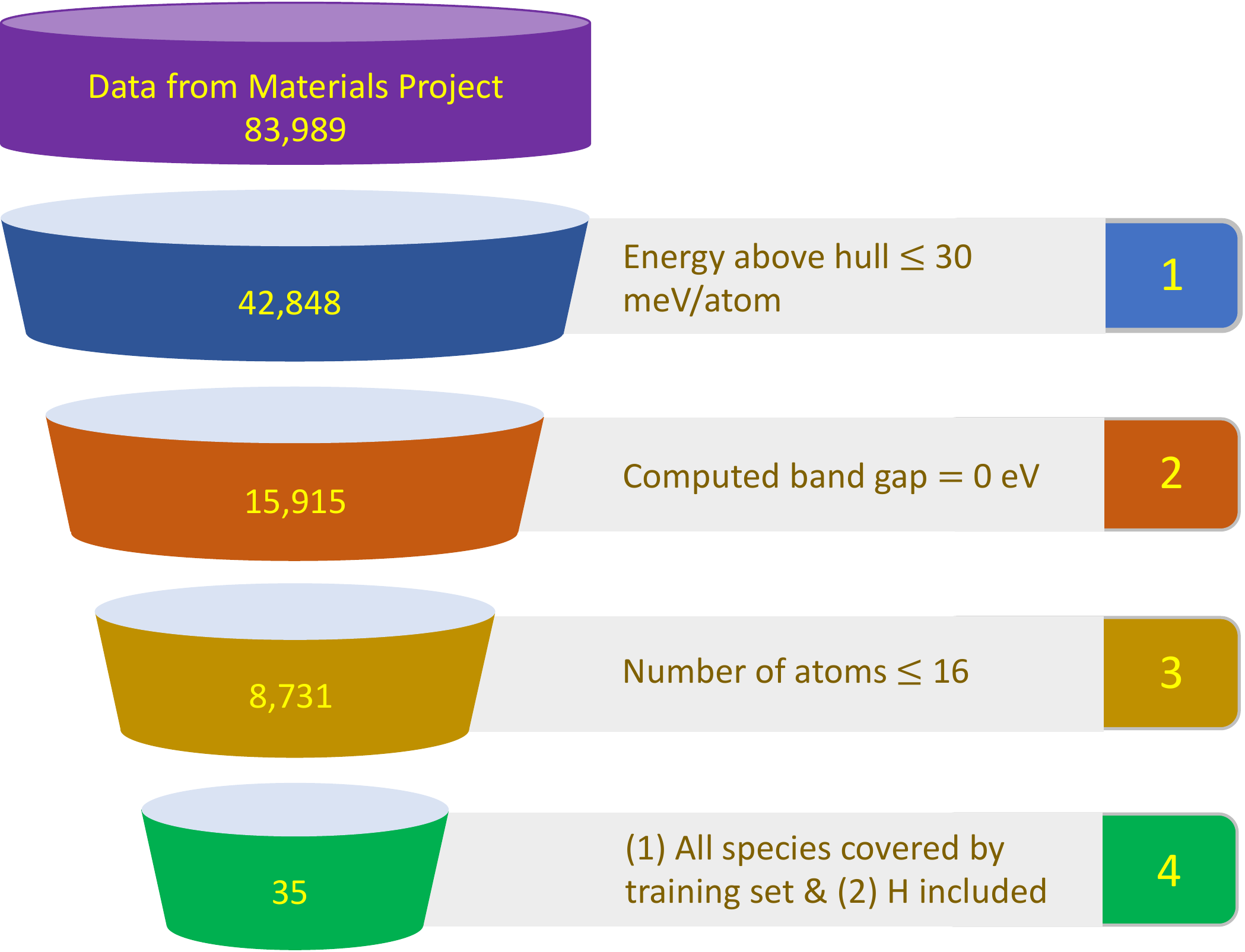}
\caption{Procedure to down select 35 atomic structures for predicting $\lambda$ and $\omega_{\log}$ from 83,989 atomic structures of Materials Project database.}\label{fig:flow2}
\end{figure}

We obtained the Materials Project database \cite{Jain2013} of 83,989 atomic structures and several properties uniformly computed at $P=0$ using {\sc vasp} \cite{vasp1, vasp3}. Starting from this dataset, we selected a subset of 35 atomic structures that have {\it energy above hull} $E_{\rm hull} < 0.03$ eV/atom, zero band gap ($E_{\rm g} = 0$ eV), no more than 16 atoms in the primitive cell, and only the species included in the training data, specifically H (see Fig. \ref{fig:data}). The first criterion ``places'' the selected atomic structures into the so-called ``amorphous limit'', a concept defined in an analysis of Materials Project database \cite{Aykoleaaq0148} and used to label the atomic structures that are (or nearly) thermodynamically stable and thus, they may be synthesized. In fact, some metastable ferroelectric phases of hafnia that are above the ground state of $\simeq 0.03$ eV/atom \cite{Huan:hafnia, Rohit:hafnia, Rohit:hafnia2} have been stabilized and synthesized \cite{Sang_hafnia, BosckeHfO2_APL2011}. Next, $E_{\rm g} = 0$ eV was used to remove non-conducting materials while the third criterion aims at selecting small enough systems for which computations of $\lambda$ and $\omega_{\log}$ are affordable. Finally, by considering only those having the species included in the training data, specifically H, we expect that the ML models will only be used in their domain of applicability. The procedure is summarized in Fig. \ref{fig:flow2}. 

The set of 35 candidates has no overlap with the training data. This set is small because the requirement of having H is very strong. In fact, removing this requirement increases the candidate set size to 2,694. Given that the ML models are extremely rapid, there is in fact no time difference between predicting $\lambda$ and $\omega_{\log}$ for 35 atomic structures and predicting these properties for 2,694 atomic structures. However, the dominance of H in the training dataset strongly suggests that the smaller set of 35 candidates is more suitable for the demonstration purpose of this work. In the next step, the training dataset will be augmented with $\lambda$ and $\omega_{\log}$ computed for materials having underrepresented species, and larger candidate sets will be examined.

\begin{figure}[t]
\centering
\includegraphics[width=1.0\linewidth]{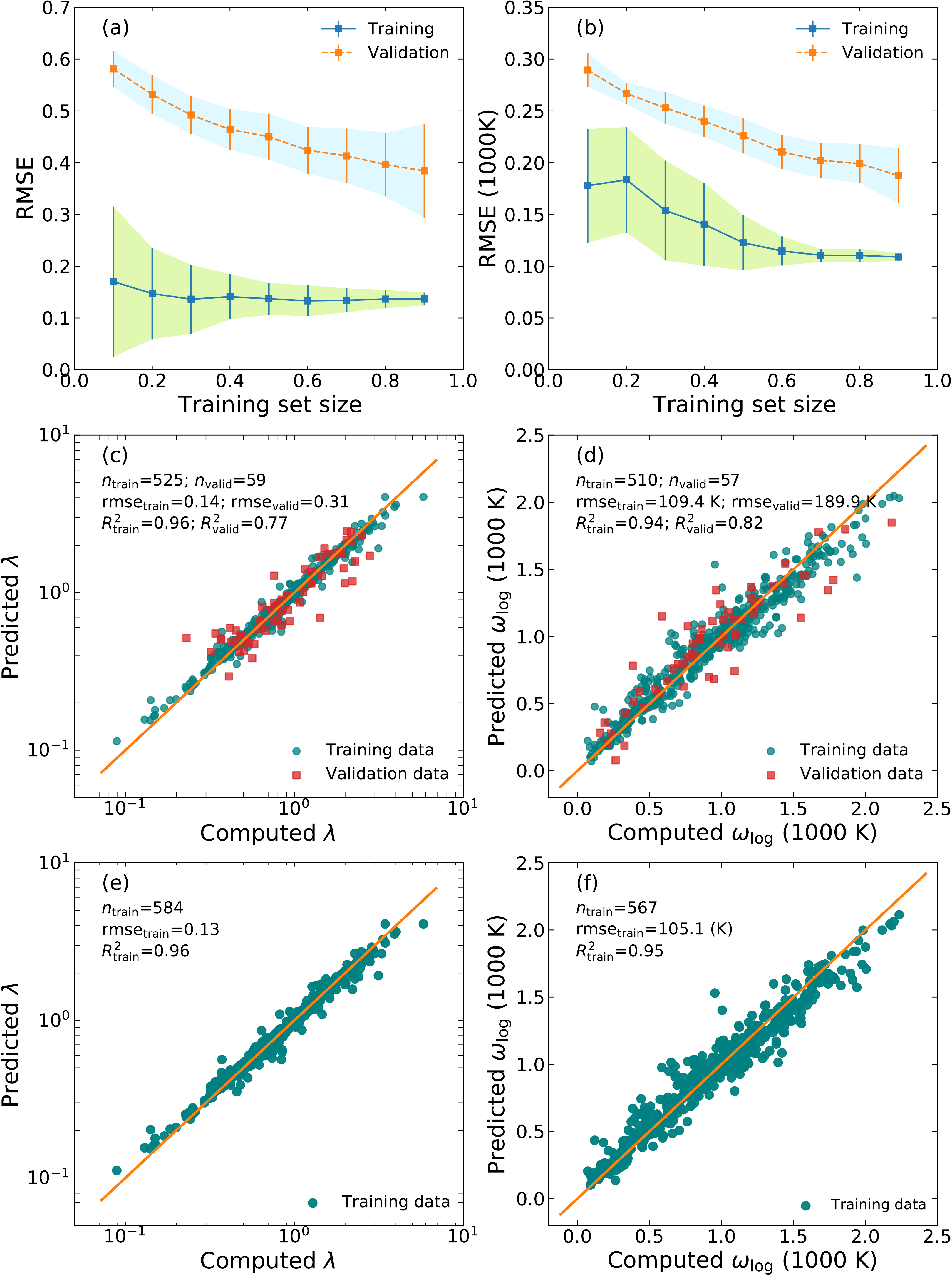}
	\caption{Learning curves obtained by learning two datasets of (a) $\lambda$ and (b) $\omega_{\log}$, a typical model trained on 90\% of the data of (c) $\lambda$ and (d) $\omega_{\log}$ and validated on the remaining {\it unseen} 10\% data, and two ML models trained on 100\% of the data of (e) $\lambda$ and (f) $\omega_{\log}$. In (a) and (b), each data point is associated with an errorbar obtained from 100 models that were independently trained.}\label{fig:model}
\end{figure}

\begin{figure}[t]
\centering
\includegraphics[width=1.0\linewidth]{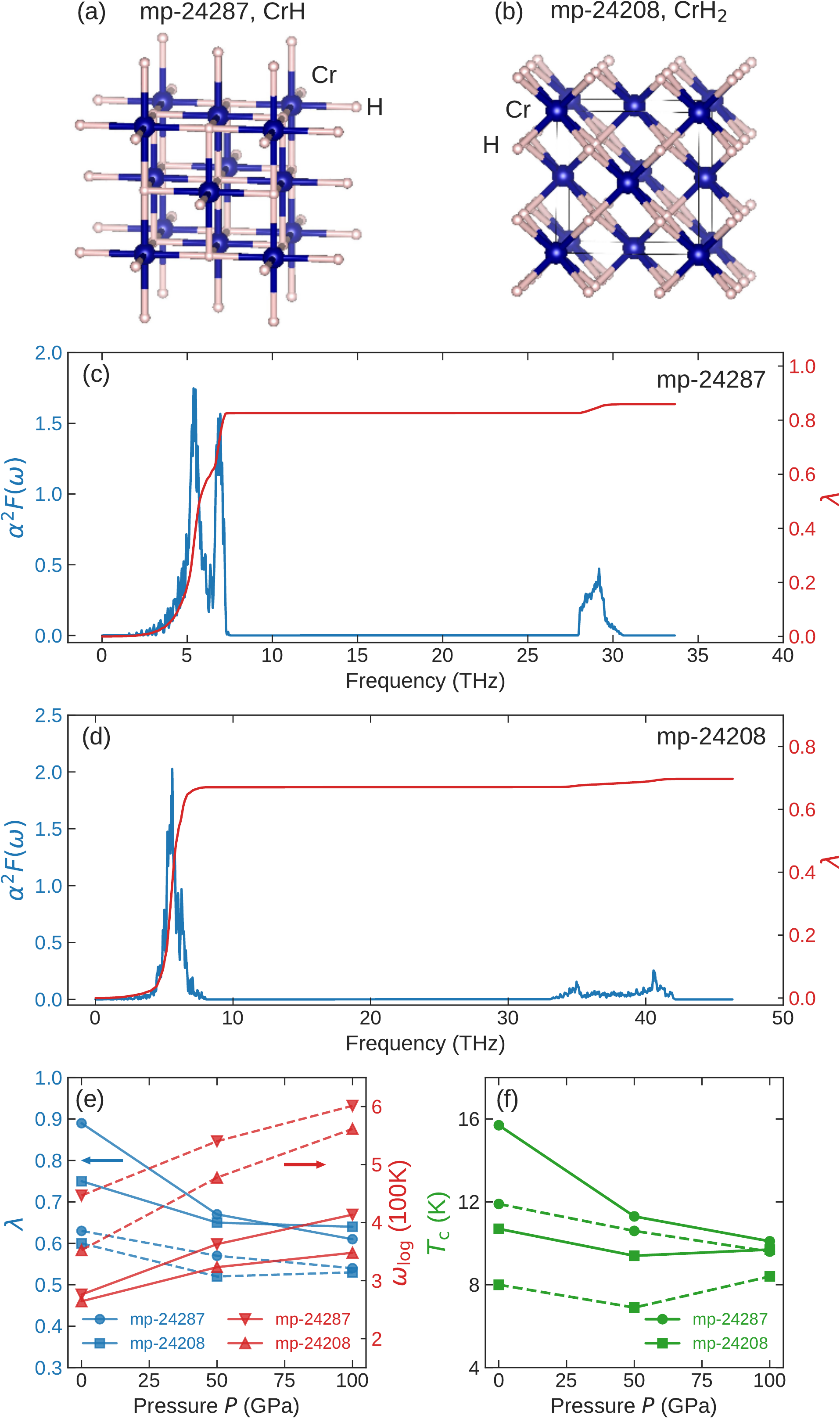}
\caption{Computed superconducting properties of mp-24287 and mp-24208, whose atomic structures are visualized in (a) and (b) and spectral function $\alpha^2F(\omega)$ and the accumulative $\lambda (\omega)$ are shown in (c) and (d). The ${\bf k}$-point and ${\bf q}$-point grids used for (c) are $24\times 24\times 24$ and $8\times 8 \times 8$, respectively, while those used for (d) are $21\times 21\times 21$ and $7\times 7 \times 7$, respectively. In (e) and (f), solid and dashed lines are used to show the computed (using the extrapolation procedure described in Sec. \ref{sec:dft}) and the predicted values $\lambda$, $\omega_{\log}$, and $T_{\rm c}$ (computed from $\lambda$ and $\omega_{\log}$ using McMillan formula with $\mu^*=0.1$) at $P = 0$, $50$, and $100$ GPa.}\label{fig:validate}
\end{figure}

\begin{table*}[t]
\begin{center}
\caption{Six hydrogen-containing materials that have highest predicted $T_{\rm c}$ among 35 materials in the candidate set, given in the top part. For each of them, the ID and the energy above hull $E_{\rm hull}$ obtained from Materials Project are given (the pressure $P$ and computed band gap are all zero). Predicted $\lambda$, $\omega_{\log}$, and $T_{\rm c}$ were obtained from the ML models and computed using McMillan formula with $\mu^*=0.1$. Among the 6 materials, computations were performed for 4 materials, two of them (mp-24287 and mp-24208) are dynamically stable, thus computed $\lambda$, $\omega_{\log}$, and $T_{\rm c}$ are available. In the bottom part of the Table, predicted and computed values of $\lambda$, $\omega_{\log}$, and $T_{\rm c}$ are reported for two dynamically stable materials, i.e., mp-24287 and mp-24208, at 50 GPa and 100 GPa.}\label{table:disc}
\begin{tabular}{lcccccccccccc}
\hline
		\hline
		\multirow{2}{*}{MP ID} &  Chemical & Space & $P$ &$E_{\rm hull}$ &\multicolumn{3}{c}{Predicted} &Computation& \multicolumn{4}{c}{Computed} \\
		\cline{10-13}
		\cline{6-8}
		 &  formula & group & (GPa)& (eV/atom) &$\lambda$&$\omega_{\rm log}$ (K)&$T_{\rm c}$ (K)& performed& Dyn. stable& $\lambda$& $\omega_{\rm log}$ (K) & $T_{\rm c}$ (K)\\
		\hline
		mp-24289 & PdH & $Fm\overline{3}m$& 0 & $0.02$ &$0.88$&$377.2$&$21.3$&Yes&No&$-$ & $-$ & $-$\\
		mp-1018133 & LiHPd & $P4/mmm$&0& 0 &$0.79$&$321.0$&$14.5$&Yes&No&$-$ & $-$ & $-$\\
		mp-24081 & ScClH & $R\overline{3}m$ &0& 0 &$0.65$&$445.9$&$13.0$&No&$-$&$-$ & $-$ & $-$\\
		mp-24287 & CrH & $Fm\overline{3}m$&0& 0 &$0.63$&$446.6$&$11.9$&Yes&Yes&$0.89$ & $276.2$ & $15.7$\\
		mp-1008376 & CeH$_3$ & $Fm\overline{3}m$&0& 0 &$0.60$&$418.5$&$9.5$&No&$-$&$-$ & $-$ & $-$\\
		mp-24208 & CrH$_2$ & $Fm\overline{3}m$ &0& 0 &$0.60$&$352.5$&$8.0$&Yes&Yes&$0.75$ & $264.4$& $10.7$\\
		\hline
		mp-24287 & CrH & $Fm\overline{3}m$&50& $-$ &  $0.57$ & $540.3$ & $10.6$ &Yes&Yes& $0.67$ & $362.8$ & $11.3$\\
		 & CrH & $Fm\overline{3}m$&100& $-$ & $0.54$ & $601.4$ & $9.6$ &Yes&Yes& $0.61$ & $413.7$ & $10.1$\\
		mp-24208 & CrH$_2$ & $Fm\overline{3}m$ &50& $-$ & $0.52$ & $477.6$ & $6.9$ &Yes&Yes& $0.65$ & $323.2$ & $9.4$\\
		 & CrH$_2$ & $Fm\overline{3}m$ &100& $-$ & $0.53$ & $561.6$ & $8.4$ &Yes&Yes& $0.64$ & $348.0$ & $9.7$\\
		\hline
		\hline
		\end{tabular}
	\end{center}
\end{table*}

\section{Results}
\subsection{Machine-learning models}\label{sec:model}
Given a learning algorithm and a dataset that has been represented appropriately, learning curves can be created using an established procedure. In this work, each dataset was randomly split into a training set and a (holdout) validation set. Next, a ML model was trained on the training set using standard 5-fold cross-validation procedure \cite{james2013introduction} to regulate the potential overfitting. Then, the ML model was tested on the validation set, which is entirely unseen to the trained model. By repeating this procedure 100 times and varying the training set size, a training curve and a validation curve  were produced from the mean and the standard deviation of the root-mean-square error (RMSE) of the predictions of the training sets and the validation sets. During the training/validating processes, randomness stems from the training/validation data splitting and the 5-fold training data splitting for internal cross validation. As such random fluctuations are suppressed statistically by averaging over 100 independent models, the learning curves could provide some useful and unbiased insights into the performance of the data, the featurize procedure, the learning algorithm, and ultimately the ML models that are developed. 

Two learning curves obtained by using GPR to learn the (featurized) $\lambda$ and $\omega_{\log}$ datasets are shown in Figs. \ref{fig:model} (a) and (b). In both cases, the training curves saturate at $\simeq 0.15$ (for $\lambda$) and $110$ K (for $\omega_{\log}$). These values are small, i.e., they are $\simeq 3-5$\% of the data range, implying that GPR can successfully capture the behaviors of the data. On the other hand, the validation curves of $\lambda$ and $\omega_{\log}$ data do not saturate and keep decreasing. This behavior strongly suggests that if more data are available, the gap between the learning and the validation curves can further be reduced and the performance of the target ML models can readily be elevated. 

Figs. \ref{fig:model} (a) and (b) reveal that an error of $\simeq 0.4$ and $\simeq 200$ K can be expected for the predictions of $\lambda$ and $\omega_{\log}$, respectively. The expected errors are roughly 7 \% of the whole range of $\lambda$ and $\omega_{\log}$ data, which are significantly small compared to the results reported in Ref. \onlinecite{choudhary2022designing}. Figs. \ref{fig:model} (c) and (d) visualize two typical ML models trained on 90\% of the $\lambda$ and $\omega_{\log}$ datasets and validated on the remaining 10\% of the datasets. Likewise, Figs. \ref{fig:model} (e) and (f) visualize two typical ML models, each of them was trained on the entire $\lambda$ or $\omega_{\log}$ dataset using the exactly same procedure. In fact, each of them is one of 100 ML models that were trained independently and used to predict $\lambda$ and $\omega_{\log}$ of the candidate set.

\subsection{Discovered superconductors and validations}\label{sec:disc}
We used the developed ML models to predict $\lambda$ and $\omega_{\log}$ of 35 atomic structures in the candidate set, and then to compute the critical temperature $T_{\rm c}$ using the McMillan formula with $\mu^*=0.1$. The predicted $\lambda$ ranges from $0.31$ to $0.88$, and consequently, the predicted $T_{\rm c}$ ranges from $0.16$ K to $21.3$ K. Six candidates with highest predicted $T_{\rm c}$ are those with Materials Project ID of mp-24289, mp-1018133, mp-24081, mp-24287, mp-1008376, and mp-24208. Details of these candidates are summarized in Table \ref{table:disc} while comprehensive information of all 35 candidates can be found in Supplemental Material \cite{supplement}. 

Examining the top six candidates, we found that mp-24081 is a trigonal structure of ScClH, whose primitive cell has 6 atoms and three very small lattice angles ($\alpha = \beta = \gamma = 21.38^\circ$). Computations of the EP interactions in such a structure are prohibitively expensive because the required $\bf k$-point and $\bf q$-point grids must be extremely large. In addition, Ce, the species showing up in mp-1008376, a cubic structure of CeH$_3$, is not supported by ONCVPSP-PBE-PDv0.4 norm-conserving pseudopotential set \cite{hamann2013optimized}. Therefore, computations were performed for the remaining four candidates. Among them, mp-24289, a cubic structure of PdH and mp-1018133, a tetragonal structure of LiHPd, are dynamically unstable. In principles, each of them can be stabilized by following the imaginary phonon modes to end up at a dynamically stable structure with lower energy and symmetry \cite{Huan:NaSc}. Such heavy and cumbersome technical procedure was reserved for the next steps. The last two candidates are mp-24287, which is a cubic structure of CrH and mp-24208, which is also a cubic structure of CrH$_2$. Both of them, visualized in Figs. \ref{fig:validate} (a) and (b), are dynamically stable and thus, their $\lambda$ and $\omega_{\log}$ were computed using the procedure described in Sec. \ref{sec:dft}. The phonon band structures, which prove the dynamical stability of mp-24289, mp-1018133, mp-24287, and mp-24208, can be found in the Supplemental Material \cite{supplement}.

Predicted and computed $\alpha^2F(\omega)$, $\lambda$, $\omega_{\log}$, and $T_{\rm c}$ (using the McMillan formula with $\mu^*=0.1$) of mp-24287 and mp-24208 at $P=0$ are given in Table \ref{table:disc} and Figs. \ref{fig:validate} (c) and (d). Considering the expected errors of the ML models, it is obvious that the computed $\lambda$ and $\omega_{\log}$ agree remarkably well with the ML predicted values. Given that magnesium diboride MgB$_2$ in its hexagonal $P6/mmm$ phase is the highest-$T_{\rm c}$ conventional superconductor with $T_{\rm c}\simeq 39$ K \cite{nagamatsu2001superconductivity}, the examined materials have respectable (computed) critical temperature, i.e., $T_{\rm c} = 15.7$ K for mp-24287 and $T_{\rm c} = 10.7$ for mp-24208. By examining the electronic structures of mp-24287 and mp-24208 reported in Materials Project database, we confirmed that both of them are metallic in nature with a large density of states at the Fermi level.

We extended our validation to the high-$P$ domain by predicting and then computing $\lambda$ and $\omega_{\log}$ of mp-24287 and mp-24208 after optimizing them at $P = 50$ GPa and $P = 100$ GPa. Both of them were found to be dynamically stable at these pressures while the computed superconducting properties are shown in Table \ref{table:disc} and Figs. \ref{fig:validate} (e) and (f). We also found that the computed and the predicted values of $\lambda$ and $\omega_{\log}$ at $P = 50$ GPa and $P = 100$ GPa are remarkably consistent. For both materials, computed $\lambda$ and $T_{\rm c}$ decrease while $\omega_{\log}$ increases from $0$ to $100$ GPa, and the ML models capture correctly these behaviors within the expected errors given from the analysis of the learning curves in Sec. \ref{sec:model}. Specifically, predictions of $\lambda$ at $P = 50$ GPa and $P = 100$ GPa are within $0.1$ from the computed results, leading to a remarkably small error of $\simeq 3$ K in predicting $T_{\rm c}$.

\begin{figure}[t]
\centering
\includegraphics[width=0.8\linewidth]{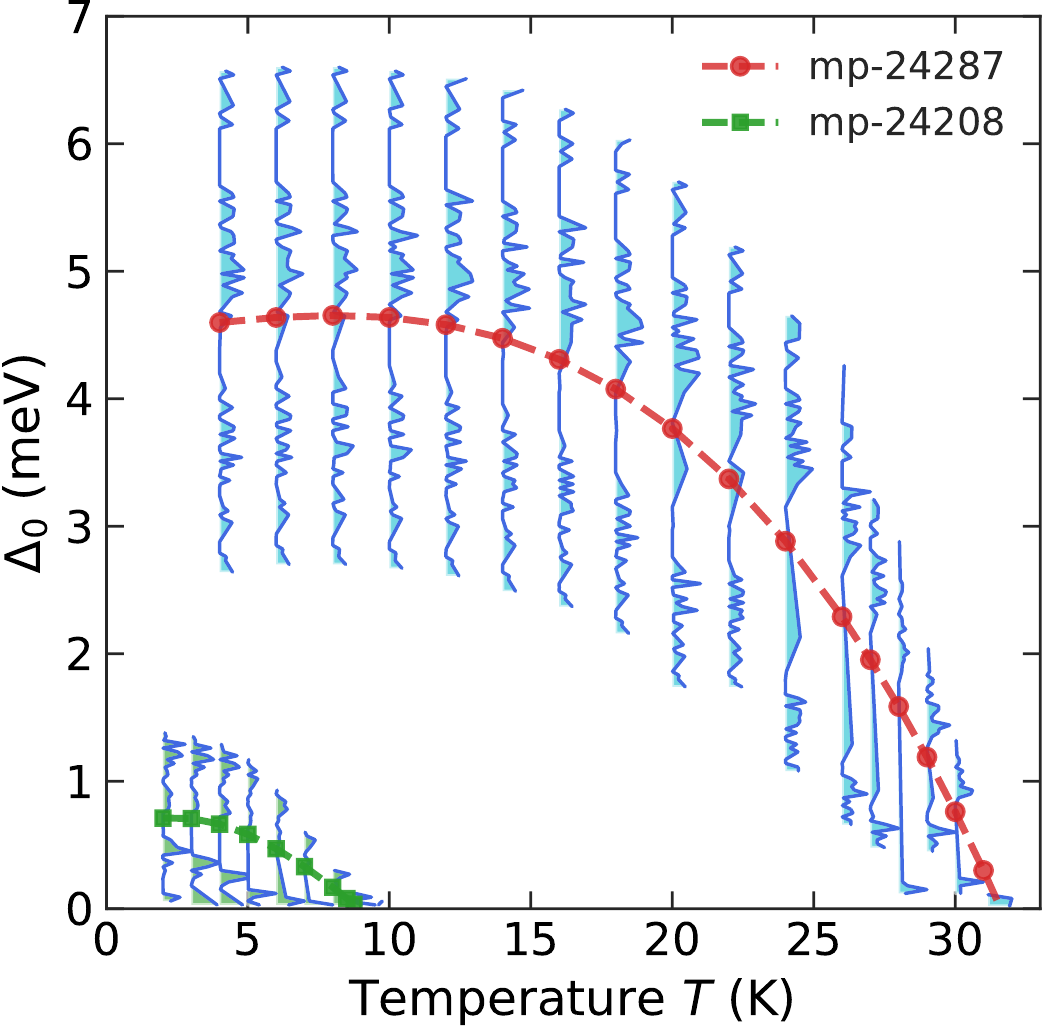}
\caption{Distribution of the zero-pressure superconducting gap function $\Delta_0 (T)$ computed by numerically solving the \'{E}liashberg equations for mp-24287 and mp-24208. The dashed curves, joining the middle point of the distributions, serve as the guide to the eyes. The critical temperature $T_{\rm c}$ is estimated to be at the middle point of the downward-sloping segment of the $\Delta_0 (T)$ curves.}\label{fig:gap}
\end{figure}

\subsection{Further assessments on the predictions}
We attempted to verify our predictions in a few ways. First, additional calculations for $\lambda$ and $\omega_{\log}$ of mp-24287 and mp-24208 using the local-density approximation (LDA) XC functional were performed at all the pressure values examined (see Sec. \ref{sec:disc}). The obtained results, as given in the Supplemental Material \cite{supplement}, are highly consistent with, i.e., within $2-3$\% of, the reported results using the PBE XC functional. 

Next, we used EPW code \cite{giustino2007electron,margine2013anisotropic, ponce2016epw} to numerically solve the \'{E}liashberg equations on the imaginary axis and then approximated the real-axis superconducting gap $\Delta_0$ of mp-24287 and mp-24208 using P\'{a}de continuation \cite{marsiglio1988iterative}. Within this scheme, the electron-phonon interactions were computed by Quantum ESPRESSO \cite{QE1, QE2}, using the ultra-soft pseudopotentials from PS Library \cite{dal2014pseudopotentials}, an energy cutoff of 120 Ry (which is 60 Ha, $\simeq 1,600$ eV), a ${\bf k}$-point grid of $24\times 24\times 24$ and a ${\bf q}$-point grid of $6\times 6 \times 6$. During the EPW calculations, we used a fine ${\bf q}$-point grid of $12\times 12 \times 12$ and $\mu^* = 0.1$. The superconducting gap $\Delta_0 (T)$ computed for mp-24287 and mp-24208 and shown in Fig. \ref{fig:gap} projects a $T_{\rm c}\simeq 22-24$ K for mp-24287 and a $T_{\rm c}\simeq 7-8$ K for mp-24208. These values are in good agreement with that reported in Fig. \ref{fig:validate}, providing a confirmation of the predicted superconductivity of mp-24287 and mp-24208 at $P=0$ GPa.

Finally, we turn our attention to the synthesizability of mp-24287 and mp-24208 by tracing their origin. Information from Materials Project database allows us to track them down to two entries numbered 191080 and 26630 of the Inorganic Crystalline Structure Database (ICSD), and finally to Refs. \onlinecite{antonov2007crystal} and \onlinecite{snavely1949unit}, respectively. In short, mp-24287 and mp-24208 were experimentally synthesized and resolved \cite{antonov2007crystal, snavely1949unit} sometimes in the past. Afterwards, some experimental \cite{pozniak2001magnetic, antonov2022lattice} and computational \cite{miwa2002first, kanagaprabha2015investigation} efforts followed, examining their magnetic, electronic, and mechanical properties. Perhaps because preparing them experimentally is challenging, little more is known about these materials. Given the documented evidence of the synthesizability of both mp-24287 and mp-24208 at $0$ GPa, which is in contrast with the enormous challenges of performing experiments at hundreds of GPa,  we hope that these materials will be resynthesized and tested for the predicted superconductivity in the near future.

\section{Remarks and going forward}
Predicting $\lambda$ and $\omega_{\log}$ from the atomic structures has some advantages. First, the correlation between the atomic structures and $\lambda$ and $\omega_{\log}$, which will be learned, is direct, physics-inspired, and intuitive, while computing $T_{\rm c}$ from  $\lambda$ and $\omega_{\log}$ is trivial. Second, the obtained ML models, which are accurate and robust, can be directly used not only for extant massive material databases with millions of atomic structures but also for any structure searches in an {\it on-the-fly} manner. Finally, by using pressure as an implicit input, the training data can be highly diverse and comprehensive, ultimately allowing the ML models to be able to handle unusual atomic environments, frequently encountered during unconstrained structure searches for new materials. 

The accuracy demonstrated in Sec. \ref{sec:disc} for the ML models of $\lambda$ and $\omega_{\log}$ is rooted from a series of factors. The list includes at least a reliable training dataset, a featurizing procedure that can capture the essential information encoded in the atomic structures, a ML algorithm that can learn the featurized data efficiently, a careful justification of the domain of applicability of the ML models, and a good candidate set. On the other hand, these stringent factors limit the number of candidates used in this work, although the ML models are already very fast to make millions of predictions. 

In the next steps, we will improve the whole scheme in several ways. First, by enlarging and diversifying the dataset while maintaining its quality, the domain of applicability of the ML models will be systematically expanded. For examples, the candidate set obtained from the selection procedure described in Fig. \ref{fig:flow2} will jump to 2,694 atomic structures when we can remove the requirement of having H in the chemical composition. Coming that point, we believe that many more new superconductors can be identified and validated, at least by first-principles computations. Second, modern deep learning techniques will be used to improve and possibly to unify the featurizing and the learning steps. Third, the ML models will be integrated in an inverse design strategy to explore the practically infinite materials space in an efficient manner. Currently, (inverse) design of functional materials with targeted properties is a very active research area with many success stories \cite{Huan:design, Arun:design, ZHANG201551,Xiang:Sidesign, fung2021inverse, coli2022inverse, lininger2021general, court2021inverse}. We hope that superconducting materials discoveries can be added to this list in the near future. Finally, we will work with experimental experts to synthesize and test the superconducting materials discovered computationally, closing the loop of materials design.

\section{Conclusions}

We have demonstrated a ML approach for the discovery of conventional superconductors at any pressure. By exploring and learning the direct and physics-inspired correlation between the atomic structures and their possible superconducting properties, specifically $\lambda$ and $\omega_{\log}$, highly accurate and reliable ML models were developed. These models were validated against the standard first-principles calculations of $\lambda$ and $\omega_{\log}$, identifying two potential superconducting materials with respectable critical temperature $T_{\rm c}$ at zero pressure. Interestingly, these materials have been synthesized and studied in some other contexts. The main implication of this approach is that by learning the high-$P$ atomic-level details that are connected to high-$T_{\rm c}$ superconductivity, the obtained ML models can be used to identify the atomic structures realized at zero pressure with possible high-$T_{\rm c}$ superconductivity. Given that the models can be used directly for massive materials databases with millions of atomic configurations, more superconductors can be expected in near future. We plan to improve this strategy in multiple ways, hoping that it can better contribute to the search of high-$T_{\rm c}$ superconductors that has been highly active during the last decade.

\section*{Acknowledgements}
Work by T.N.V. was supported by Vingroup Innovation Foundation (VINIF) in project code VINIF.2019.DA03. The authors thank Chris Pickard, Guochun Yang, Bin Li, Samuel Ponc\'{e}, and Kamal Choudhary for useful communications. Computations were performed at the San Diego Supercomputer Center (Expanse) within the XSEDE/ACCESS allocation number DMR170031.


%
\end{document}